\documentclass[
reprint,
superscriptaddress,
 amsmath,amssymb,
aps,
]{revtex4-2}

\usepackage{color}
\usepackage{graphicx, physics}
\usepackage{dcolumn}
\usepackage{bm}
\usepackage{float}
\usepackage[breaklinks,colorlinks=true,linkcolor=blue,urlcolor=blue,citecolor=blue]{hyperref}

\begin{document}


\title{Current-control of chaos and effects of thermal fluctuations in magnetic tunnel junctions}

\author{Ryo Tatsumi}
 \email{tatsumi.ryo.p6@dc.tohoku.ac.jp}
 \affiliation{Department of Applied Physics, Graduate School of Engineering, Tohoku University, Sendai, Miyagi 980-8579, Japan}
 \affiliation{Department of Information Science and Technology, Graduate School of Science and Engineering, Yamagata University, Yonezawa, Yamagata 992-8510, Japan}
\author{Shinji Miwa}
 \affiliation{The Institute for Solid State Physics, The University of Tokyo, Kashiwa, Chiba 277-8581, Japan}
 \affiliation{Trans-scale Quantum Science Institute, The University of Tokyo, Bunkyo, Tokyo 113-0033, Japan}
\author{Hiroaki Matsueda}
 \affiliation{Department of Applied Physics, Graduate School of Engineering, Tohoku University, Sendai, Miyagi 980-8579, Japan}
 \affiliation{Center for Science and Innovation in Spintronics, Tohoku University, Sendai 980-8577, Japan}
\author{Takahiro Chiba}
 \email{t.chiba@yz.yamagata-u.ac.jp}
 \affiliation{Department of Applied Physics, Graduate School of Engineering, Tohoku University, Sendai, Miyagi 980-8579, Japan}
 \affiliation{Department of Information Science and Technology, Graduate School of Science and Engineering, Yamagata University, Yonezawa, Yamagata 992-8510, Japan} 


\date{\today}
 
\begin{abstract}
We theoretically investigate the chaotic behavior of spin-torque ferromagnetic resonance in magnetic tunnel junctions (MTJs) with perpendicular magnetic anisotropy under thermal fluctuations. 
By calculating the Lyapunov exponent based on the Landau–Lifshitz–Gilbert equation, we demonstrate that an MTJ characterized by a double-well potential, composed of uniaxial magnetic anisotropy and an external magnetic field, exhibits chaotic magnetization dynamics that can be controlled by means of the DC current bias. 
Furthermore, we find that thermal fluctuations help to induce these chaotic magnetization dynamics, which can be regarded as noise-induced chaos.
This research provides a basis for brain-inspired computing using spintronic devices and advances the understanding of the interplay between thermal fluctuations and chaos in magnetization dynamics.
\end{abstract}


\maketitle

{\it{Introduction.}}
In recent years, there has been growing interest in the nonlinear and complex dynamics of physical systems for brain-inspired computing, including chaotic neural networks \cite{Aihira90, Yamada93, Pan20}, reservoir computing \cite{Tanaka19, Paquot12, Torrejon17}, and probabilistic computing \cite{Bhatti17, Camsari19, Lee25}. 
As demonstrated in previous studies, it is often reported that reservoir computing achieves high computational performance when their elements exhibit ``edge of chaos'', which is a transient state between the periodic and chaotic states \cite{Aihira90, Legenstein07, Boedecker12, Bertschinger04}. 
Owing to the rich nonlinear effects and fast response of magnetization, spintronic devices have been actively implemented in experiments for the brain-inspired computing \cite{Torrejon17, Bhatti17, Furuta18, Camsari19}.
At the same time, theoretical studies of nonlinear and chaotic dynamics in spintronic devices have been actively conducted across various platforms, including ferromagnetic resonance (FMR) \cite{Mayergoyz09, Tatsumi25, Chiba26}, spin waves \cite{Wang11, Ustinov21}, magnetic-vortex-based devices \cite{Petit12, Horizumi25}, and spin-torque oscillators \cite{Li06, Yamaguchi19, Taniguchi19}. 

Among these studies, the magnetic Duffing oscillator is proposed \cite{Tatsumi25, Chiba26}, which exhibits chaotic magnetization dynamics controlled by DC and AC magnetic fields.
The magnetic Duffing oscillator is characterized by the homoclinic orbit\cite{Wiggins03}, which is the origin of chaos, in the phase space of magnetization dynamics.
The homoclinic orbit resides in a double-well potential composed of uniaxial magnetic anisotropy and an external DC magnetic field.
By tuning the DC magnetic field applied perpendicular to the magnetic anisotropy axis, one can shift the position of the homoclinic orbit, adjusting the onset of chaos.
Although the magnetic Duffing oscillator has substantial practical value on spintronic devices, it requires the adjustment of both AC and DC magnetic fields for the generation and control of chaos, which makes it less suitable for integration into nanodevices.

Magnetic tunnel junctions (MTJs) \cite{Yuasa04, Zhu06} enable electrical input and readout of the magnetization state without relying on additional magnetic sensors, making them highly suitable for miniaturization and integration. 
These characteristics have attracted considerable attention to their magnetization dynamics\cite{Tulapurkar05,Sankey06,Kubota13,Tsujikawa09, Iihama14, Lattery18}.
Since MTJs are primarily designed for room temperature operation, it is essential to consider the influence of thermal fluctuations, that is, stochastic forces, when investigating their nonlinear dynamics. 
Up to now, nonlinear dynamics under stochastic forces have been extensively studied in various physical systems \cite{Benzi81, Gammaitoni98, Pikovsky97,Bulsara90, Gan06,Lei17,Gassmann97}.
These studies report a wide range of phenomena, including stochastic resonance \cite{Benzi81, Gammaitoni98, Pikovsky97}, noise-induced chaos \cite{Bulsara90,Gan06,Lei17}, and noise-induced order \cite{Gassmann97}. 
When a stochastic force is introduced into chaotic systems, the stability of chaos becomes a highly nontrivial issue.
Nevertheless, from the perspective of utilizing strong nonlinear or chaotic dynamics in real devices, it is crucial that these behaviors remain robust against noise.

In this Letter, we theoretically investigate nonlinear and chaotic magnetization dynamics in spin-torque FMR \cite{Tulapurkar05,Sankey06} based on an MTJ with perpendicular magnetic anisotropy \cite{Tsujikawa09,Kubota13,Iihama14, Lattery18}. 
By applying an external DC magnetic field along the in-plane direction of the free layer, we create a potential landscape analogous to that of the magnetic Duffing oscillator. 
When an AC current is applied to drive the magnetization, the emergence of chaos is confirmed through the Lyapunov exponent \cite{Pikovsky16, Shimada79} and the magnetization trajectory in the phase space.
We then examine how the magnitudes of the DC and AC currents affect the onset of chaos via the spin-transfer torque. 
In particular, we find that the DC current acts as an effective control parameter which can suppress chaotic behavior. 
Furthermore, we take into account thermal fluctuations as stochastic magnetic fields,  demonstrating that they help to excite the magnetization and to enhance chaotic dynamics.

\begin{figure}[tb]
\begin{centering}
\includegraphics[width=0.45\textwidth,angle=0]{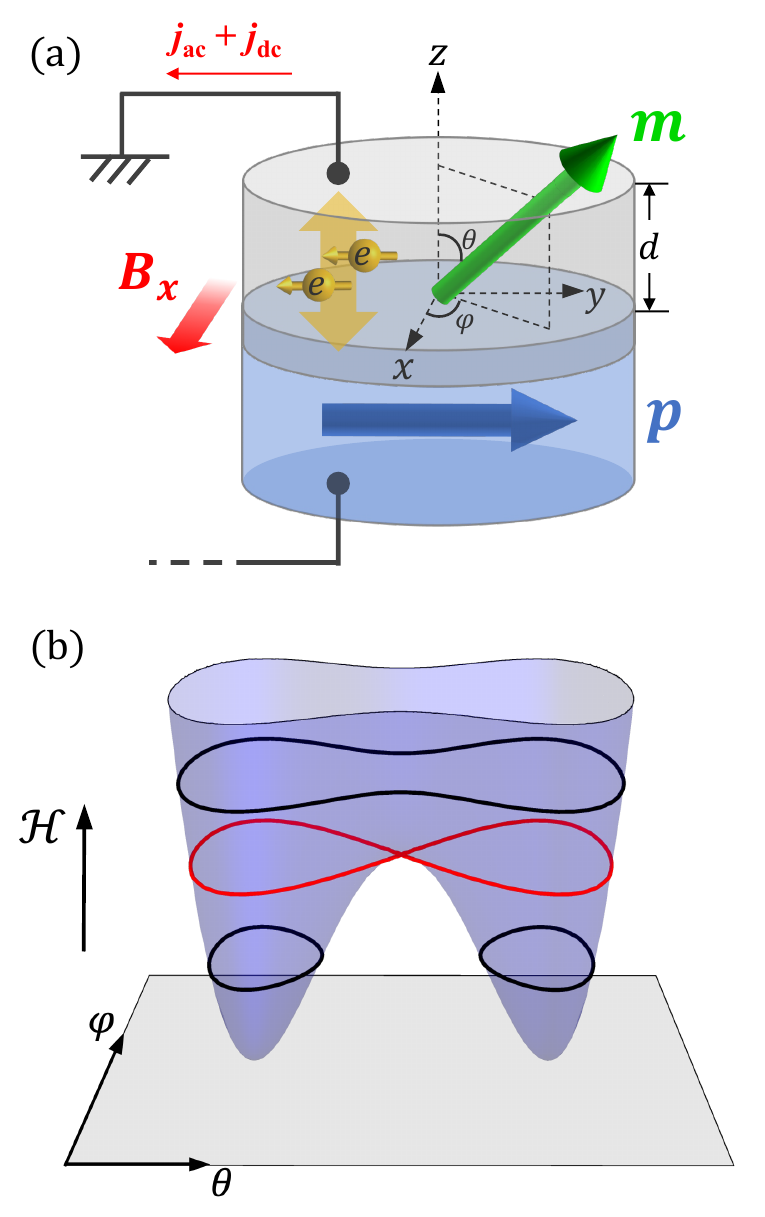} 
\par\end{centering}
\caption{
(a) Schematic illustration of the system.
The MTJ consists of a free layer and a reference layer underneath.
An external DC magnetic field $B_x$ is applied along the $x$-axis and the uniaxial magnetic anisotropy field $B_K$ is oriented along the $z$-axis.
The polar and azimuth angles of the magnetization vector in the free layer, $\bm{m}$, are denoted by $\theta$ and $\varphi$.
The magnetization vector pointing along the $y$-axis in the reference layer is represented by $\bm{p}$.
The applied AC and DC currents along the $z$ direction generate spin-transfer torque.
(b) Hamiltonian ($\mathcal{H}$) landscape of the MTJ for $B_x = 32$~mT and $B_\mathrm{ani} = 41.9$~mT.
The red energy contour line is the homoclinic orbits, which starts from and end at the same saddle point.}
\label{fig:model}
\end{figure}

{\it Model and mechanism of chaos.}
We introduce the system: an MTJ consisting of a magnetized free layer with an easy uniaxial z-axis and a magnetized reference layer in the plane, as shown in Fig.~\ref{fig:model}.
The unit vectors, denoted by $\bm{m}$ and $\bm{p}$, represent the magnetization in the free and reference layers, respectively.
The magnetization in the reference layer is pinned along the $y$-axis and induces a spin-transfer torque when an electric current is applied along the $z$-direction.
Magnetization dynamics in the free layer is described by the Landau–Lifshitz–Gilbert (LLG) equation: 
\begin{equation}
\begin{split}
    \frac{d \bm{m}}{dt} =& - \gamma \bm{m} \times \bm{B}_{\mathrm{eff}} - \gamma \tau(\bm{m}, j) \bm{m} \times (\bm{p} \times \bm{m})\\
    & \qquad +\alpha \bm{m} \times \frac{d \bm{m}}{dt},
\end{split}
    \label{LLG equation}
\end{equation}
where $\gamma$ is the gyromagnetic ratio and $\alpha$ is the Gilbert damping constant.
Then, the Hamiltonian is described as
\begin{equation}
    \mathcal{H} = -\gamma M_s \left( B_x m_x +\frac{1}{2} B_{\mathrm{ani}}m_z^2 \right),
\end{equation}
and the effective magnetic field is
\begin{equation}
\begin{split}
    &\bm{B}_{\mathrm{eff}} = - \frac{1}{M_s}\frac{\delta \mathcal{H}}{\delta \bm{m}}  = B_x \bm{e}_x + B_\mathrm{ani}m_z \bm{e}_z
\end{split}
\label{magnetic_field}
\end{equation}
where $\{{\bm{e}_x,\bm{e}_y,\bm{e}_z}\}$ are the unit vectors along the respective Cartesian axes.
$B_x$ is the DC magnetic field applied along the $x$-axis and $B_\mathrm{ani} = B_\mathrm{K} - \mu_0M_\mathrm{s}$ is the total magnetic anisotropy field consisting of $B_\mathrm{K}$ and $\mu_0M_\mathrm{s}$ ($\mu_0$ is the permeability of free space) being the interfacial\cite{Tsujikawa09,Kubota13,Iihama14, Lattery18} and shape anisotropy fields, respectively.
The second term of the right hand side of Eq.~\eqref{LLG equation} describes the spin-transfer torque which has the factor \cite{Yamaguchi19}
\begin{equation}
    \tau(\bm{m}, j) = \frac{\hbar j} {2e M_\mathrm{s} d}\frac{\eta}{1 + \eta^2 \bm{m} \cdot \bm{p}},
\end{equation}
where $\hbar$, $e$, $d$, $j$ and $\eta$ are the Dirac constant, elementary charge, the thickness of the free layer, the interfacial current density through $z$-axis, and the spin polarization of the current, respectively.
The interfacial current density $j$ is composed of two parts
\begin{equation}
    j = j_{\mathrm{dc}} + j_{\mathrm{ac}} \cos(2\pi f t),
\end{equation}
where $j_{\mathrm{dc}}$ is the interfacial DC current density, and $j_{\mathrm{ac}}$ and $f$ denote the amplitude and frequency of the interfacial AC current density, respectively.
Due to the factor $\tau$, the auto-oscillation of the magnetization and the corresponding limit cycle in the phase space can appear in the presence of the DC current.
Throughout this paper, we use the following parameters: $\gamma = 1.76 \times 10^{11}$~$\mathrm{T^{-1} s^{-1}}$, $d = 2\ \mathrm{nm}$, $\eta = 0.537$, $\mu_0M_\mathrm{s} = 1.821$~T, $\alpha = 0.05$, and $B_\mathrm{K} = 1.862$~T. 
These values represent typical parameters for an MTJ with a perpendicularly magnetized free layer consisting of CoFeB/MgO structure, with a few nanometers thickness \cite{Yamaguchi19,Kubota13}.
The fundamental principle of the giant value of $B_{\mathrm{K}}$ is broadly considered to be the hybridization of Fe-3d and O-2p orbitals at the metal-insulator interface.

In order to calculate the magnetization dynamics and the Lyapunov exponent, we express Eq.~(\ref{LLG equation}) in polar coordinates as
\begin{equation}
\begin{split}
    \frac{d \theta}{dt} &= \gamma B_\varphi(\theta, \varphi, t)
    + \alpha \gamma B_\theta(\theta, \varphi, t)
    + \gamma \tau_\theta(\theta, \varphi, z),\\
    \frac{d \varphi}{dt} &= \frac{1}{\sin{\theta}} \left[- \gamma B_\theta(\theta, \varphi, t)
    + \alpha \gamma B_\varphi(\theta, \varphi, t)
    + \gamma \tau_\varphi(\theta, \varphi, z)\right],\\
    \frac{d z}{dt} &= 2 \pi f,
\end{split}
\label{polar_LLG_full}
\end{equation}
where $B_{\theta}$ and $B_{\varphi}$ denote the $\theta$ and $\varphi$ components of the effective magnetic field $\bm{B}_{\mathrm{eff}}$, respectively, and $\tau_\theta$ and $\tau_\varphi$ represent the corresponding components of the spin-transfer torque term.
\begin{equation}
\begin{split}
    B_\theta(\theta, \varphi, t) &= (B_x + B_x^{\mathrm{th}}(t)) \cos{\theta} \cos{\varphi} \\
    & \qquad \ \ + (B_\mathrm{ani} \cos{\theta} + B_y^{\mathrm{th}}(t)) \sin{\theta} - B_z^{\mathrm{th}}(t) \sin{\theta},\\
    B_\varphi(\theta, \varphi, t) &= - (B_x + B_x^{\mathrm{th}}(t))\sin{\varphi} +B_y^{\mathrm{th}}(t) \cos{\varphi},\\
    \tau_\theta(\theta, \varphi, z) &= \tau(\theta, \varphi, z) \cos{\theta} \sin{\varphi},\\
    \tau_\varphi(\theta, \varphi, z) &= \tau(\theta, \varphi, z) \cos{\varphi}.
\end{split}
\end{equation}
where $B_i^{\mathrm{th}}(t)$ denotes the $i$-th component of the Gaussian random magnetic field, which represents the thermal fluctuation.
$B_i^{\mathrm{th}}(t)$ satisfies $\langle B_i^{\mathrm{th}}(t) \rangle = 0$ and $\langle B_i^{\mathrm{th}}(t) B_j^{\mathrm{th}}(s) \rangle = \sigma^2 \delta_{ij} \delta(t-s)$ \cite{Koch04,Taniguchi12}, where
\begin{equation}
    \sigma = \sqrt{\frac{2 \alpha k_\mathrm{B} T}{\gamma M_s \mathcal{V}}},
\end{equation}
$k_B$, $T$, $\delta_{ij}$, and $\delta(t)$ are the Boltzmann constant, the system temperature, Kronecker's deltas, and Dirac's deltas, respectively.
The volume of the free layer $\mathcal{V}$ is assumed $\pi \times 120 \times 120 \times 2\ \mathrm{nm^3}$.
We simulated the magnetization dynamics by using the Runge-Kutta method with the DifferentionalEquation package in Julia 1.10.

The mechanism of chaos, as in the magnetic Duffing oscillator, can be understood from the homoclinic orbit created by the anharmonic magnetic potential \cite{Tatsumi25,Chiba26}.
Owing to the uniaxial magnetic anisotropy field and the DC magnetic field, a double-well potential appears in the $x$–$z$ plane (i.e., along the $\theta$-axis), whereas a single-well potential emerges along the $y$-axis (see Fig.~\ref{fig:model}(b)).
The saddle exists at $(\theta, \varphi) = (\pi/2, 0)$.
Then, the orbits starting from and ending at this saddle point arise; these orbits correspond to the homoclinic orbits.
Both the Duffing oscillator and the magnetic Duffing oscillator exhibit chaotic behavior when the state is exited to the strange attractor created by the perturbed homoclinic orbit due to a driving force and damping term.
Thus, our strategy for generating or controlling chaos in the MTJ is as follows: by applying an AC current, we excite the magnetization up to the homoclinic orbit and create a strange attractor around this orbit.
By adjusting the DC current, we shift the magnetization state to modify its distance from this orbit and collapse the strange attractor.

\begin{figure}[hptb]
\begin{centering}
\includegraphics[width=0.5\textwidth,angle=0]{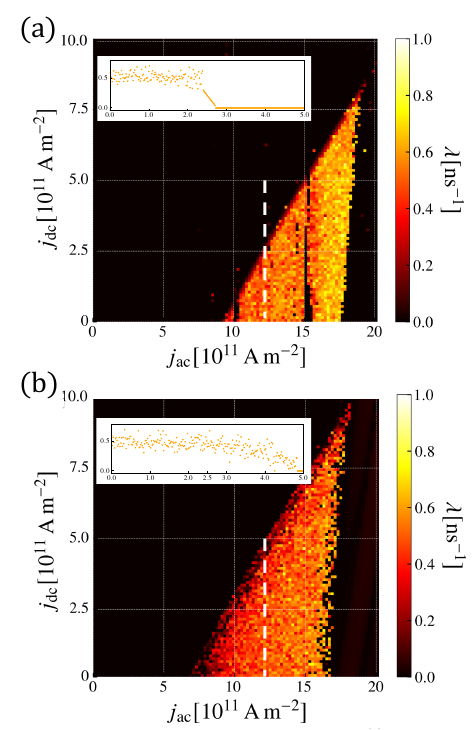} 
\par\end{centering}
\caption{
Heatmaps of the Lyapunov exponent $\lambda$ as functions of the DC current density $j_\mathrm{dc}$ and the amplitude of AC current density $j_\mathrm{ac}$ for (a) $\sigma = 0$ and (b) $\sigma = 0.366$ (corresponding to room temperature).
}
\label{fig:Lya_maps}
\end{figure}

{\it Current-control of chaos and effects of thermal fluctuations.} 

\begin{figure*}[hptb]
\begin{centering}
\includegraphics[width=1.0\textwidth,angle=0]{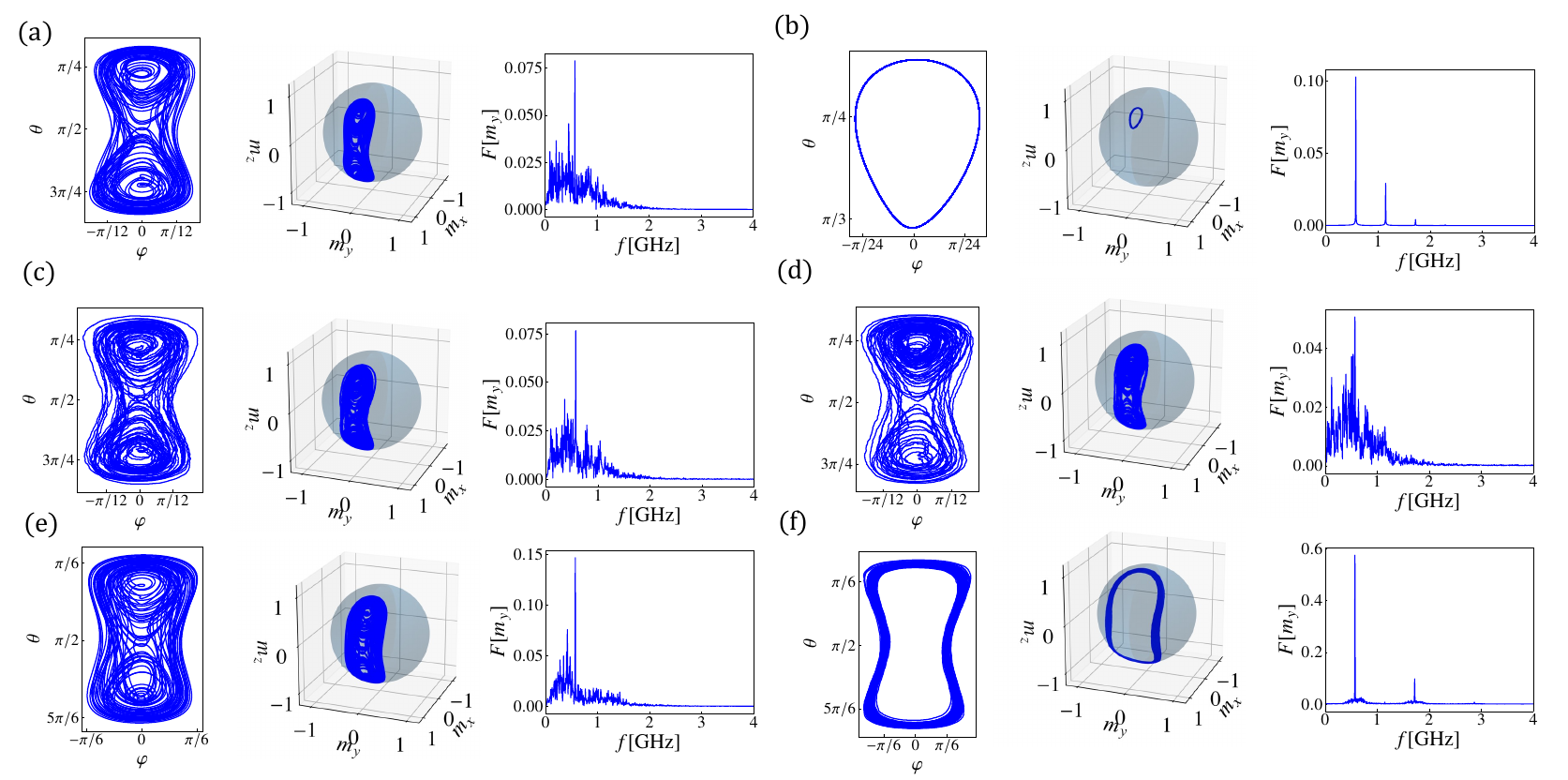} 
\par\end{centering}
\caption{
Magnetization dynamics and corresponding Fourier spectrum.
The magnetization trajectory in the phase space of $\theta$ and $\varphi$ is shown in the left panel.
The central panel displays the trajectory in three dimensional real space represented by $(m_x, m_y, m_z)$. 
The right panel shows the Fourier spectrum of the $y$-component of the magnetization $m_y$. 
The parameters are as follows: (a), (c) $j_{\mathrm{ac}} = 1.2 \times 10^{12}$~$\mathrm{Am^{-2}}$, $j_{\mathrm{dc}} = 0$; 
(b), (d) $j_{\mathrm{ac}} = 1.2 \times 10^{12}$~$\mathrm{Am^{-2}}$, $j_{\mathrm{dc}} = 2.9 \times 10^{11}$~$\mathrm{Am^{-2}}$; and 
(e), (f) $j_{\mathrm{ac}} = 1.75 \times 10^{12}$~$\mathrm{Am^{-2}}$, $j_{\mathrm{dc}} = 2.9 \times 10^{11}$~$\mathrm{Am^{-2}}$, respectively. 
Panels (a), (b) and (e) show the results without thermal fluctuations ($\sigma = 0$), whereas panels (c), (d) and (f) include thermal fluctuations at room temperature ($\sigma = 0.366$).
}
\label{fig:phase}
\end{figure*}

To confirm the presence of chaos and clarify the effect of the DC current on the magnetization dynamics, we numerically calculate the Lyapunov exponent.
Figure~\ref{fig:Lya_maps}(a) shows the Lyapunov exponent as a function of the AC current density $j_{\mathrm{ac}}$ with $f = 0.57 \ \mathrm{GHz}$ and the DC current density $j_{\mathrm{dc}}$ without thermal fluctuations.
The frequency $f$ is chosen as the one that most effectively promotes chaos \cite{SuppMat}.
In this figure, regions with $\lambda > 0$ (bright areas) indicate chaotic magnetization dynamics and $\lambda = 0$ corresponds to the limit-cycle.
As expected, when an AC current is applied, the magnetization exhibits chaotic dynamics. 
In fact, as shown in Fig.~\ref{fig:phase}(a), the AC current excites the magnetization state from the bottom of the potential well.
On the other hand, the DC current increases the threshold amplitude of the AC current required for the onset of chaotic dynamics in Fig.~\ref{fig:Lya_maps}(a). 
Note that some dark regions with a Lyapunov exponent $\lambda = 0$ appear between chaotic domains, for example, around $(j_{\mathrm{ac}},j_{\mathrm{dc}}) = (15,0)$
This phenomenon is called "periodic window" \cite{Strogatz24} and is observed in many dynamical systems that exhibit chaos.
A clear boundary between chaos and limit-cycle regimes is observed in Fig.~\ref{fig:Lya_maps}(a).

Figure~\ref{fig:Lya_maps}(b) shows the heatmap of the Lyapunov exponent with random magnetic fields at room temperature i.e., $T=300$.
Comparing Fig.~\ref{fig:Lya_maps}(a) and (b), the boundary becomes less distinct and shifts to the left due to the thermal fluctuations.
The shift in the parameter range of the chaotic region indicates not only that the strange attractor are robust against stochastic forces, but also that stochastic forces assist in exciting the magnetization, leading to chaotic dynamics.
Since the boundary of the bright region is blurred, the “edge of chaos” may also become indistinct.
Note that several studies on Duffing oscillators with stochastic forces have reported that stochastic forces can facilitate the onset of chaotic dynamics \cite{Gan06,Lei17}.
Thus, the MTJ exhibits behavior similar to that of the Duffing oscillator when subjected to stochastic forces.

Next, we examine the roles of the DC current and stochastic forces using the magnetization trajectory.
Figure~\ref{fig:phase} shows the magnetization dynamics in the phase space and the Fourier spectrum of the $y$-component of magnetization $m_y$, which can be detected by the tunnel magnetoresistance effect via the spin-torque diode effect\cite{Yuasa04,Zhu06,Tulapurkar05}.
Figure~\ref{fig:phase}(a) presents the chaotic dynamics over the double-well potential driven by an AC current density $j_{\mathrm{ac}} = 1.2 \times 10^{12}$ $ \mathrm{Am^{-2}}$ without a DC current.
This trajectory covers a wide area of the phase space, and the Fourier spectrum is broad.
When a DC current is applied, a limit cycle appears accompanying several Fourier peaks and the chaotic dynamics disappear in the phase space, as shown in Fig.~\ref{fig:phase}(b).
This is because the DC current generates a spin-transfer torque along the $y$-axis, which causes the magnetization to remain within one side of the double-well potential. 
Hence, due to the DC current, it is difficult to touch the homoclinic orbits and chaotic dynamics are suppressed.
In terms of the dynamical systems theory, the DC current acts as a bifurcation parameter that alters the qualitative structure of attractors in the phase space, such as the limit cycles or strange attractor.
Figure~\ref{fig:phase}(c) shows the magnetization dynamics with stochastic forces at room temperature.
As seen in Fig.~\ref{fig:Lya_maps} and \ref{fig:phase}, the chaotic dynamics are robust against the stochastic forces, indicating the feasibility of observing chaos experimentally.
Moreover, when stochastic forces are applied to a state in which chaos is not originally present, a strange attractor-like structure emerges, as shown in Fig.~\ref{fig:phase} (d), leading to the onset of chaotic dynamics.
This phenomenon is known as noise-induced chaos and has been reported in studies of the Duffing oscillator \cite{Gan06,Lei17}.
Physically, stochastic forces can be regarded as an additional energy source that pushes the magnetization state toward the homoclinic orbit in the phase space, leading to the emergence of chaotic dynamics. 
That is why the threshold of chaos is reduced in the presence of thermal fluctuations.

Also, we demonstrate noise-induced order, a phenomenon in which stochastic forces suppress the characteristic features of chaos, such as strange attractors and high sensitivity to initial conditions.
Figure \ref{fig:phase}(e) displays the chaotic dynamics for $j_{\mathrm{ac}} = 1.75 \times 10^{12}$ $ \mathrm{A m^{-2}}$ and $j_{\mathrm{dc}} = 2.9 \times 10^{11}$ $ \mathrm{A m^{-2}}$. 
When thermal fluctuations are introduced, the trajectories no longer wander erratically through the phase space; instead, blurred periodic trajectories emerge in Fig.~\ref{fig:phase}(f).
The Fourier spectrum contains two main peaks and sidebands around them.
In this model, the energy contours are directly related to the homoclinic orbits.
Hence, this phenomenon can be interpreted by considering thermal fluctuations as an effective energy source.
The stochastic forces push the magnetization state above the homoclinic orbits and prevent the emergence of a strange attractor around these orbits.
Thus, analyzing the occurrence of chaos through the homoclinic orbit clarifies the physical roles of terms such as spin torque or random magnetic field in generating chaotic behavior. 
In this way, the MTJ model can experimentally demonstrate not only simple chaotic magnetization dynamics but also various noise-induced phenomena known in the field of dynamical systems.

{\it Stochastically driven chaos.}
\begin{figure}[t]
\begin{centering}
\includegraphics[width=0.5\textwidth,angle=0]{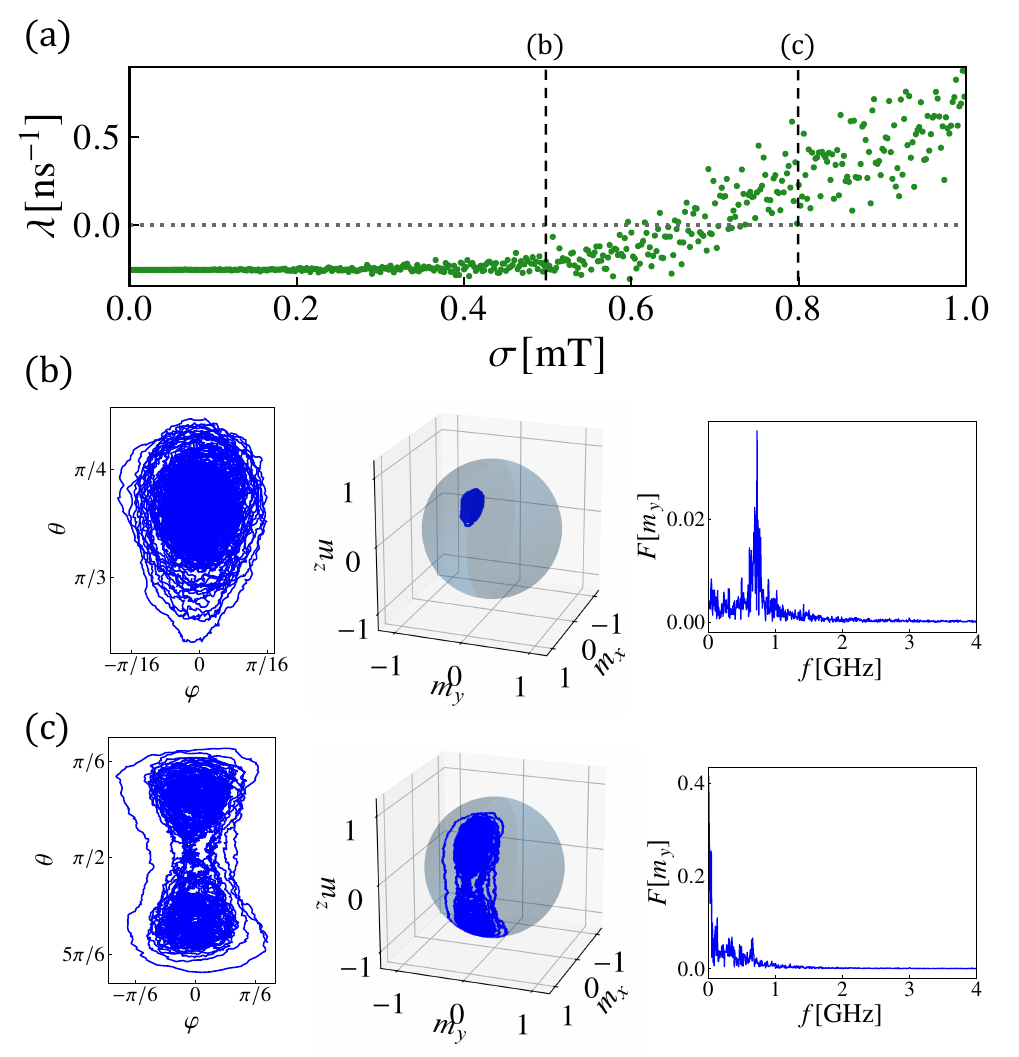} 
\par\end{centering}
\caption{
Magnetization dynamics driven solely by stochastic forces.
(a) Lyapunov exponent $\lambda$ as a function of the noise strength $\sigma$. The dynamics exhibits chaotic behaviors for $\sigma \gtrsim 0.6$.
(b),(c) Magnetization trajectory and the corresponding Fourier spectrum at (b) $\sigma = 0.5$ for the radius of the free layer being $88$~$\mathrm{nm}$ and (c) $\sigma = 0.8$ for the radius with $55$~$\mathrm{nm}$.}
\label{fig:driven_thermal}
\end{figure}
Here, we demonstrate chaotic magnetization dynamics driven solely by thermal fluctuations.
The method for evaluating the Lyapunov exponent differs from Fig.\ref{fig:Lya_maps} because the system is a two-dimensional autonomous system with a stochastic force.
We use the norm defined in the phase space $(\theta,\varphi) \in \mathbb{R}^2$.
It is noted that the stochastic force can be considered to be the third dimension in the phase space, which allows for the emergence of chaos.
Figure~\ref{fig:driven_thermal}(a) illustrates the Lyapunov exponent as a function of the standard deviation of the stochastic force $\sigma$.
As $\sigma$ increases, the Lyapunov exponent also increases and becomes positive around $\sigma = 0.6$. 
This indicates that under small stochastic forces, the trajectory is attracted to the minimum of the potential, and the system remains stable.
According to Fig.~\ref{fig:driven_thermal}(b), the magnetization is trapped on one side of the potential well.
In contrast, when $\sigma = 0.8$, the magnetization reverses between double-well potential repeatedly in Fig.~\ref{fig:driven_thermal}(c), and the Lyapunov exponent becomes positive.
This shows that the homoclinic orbit imparts the feature of chaos, sensitivity to initial states, to the magnetization dynamics driven solely by the stochastic force. 
This phenomenon indicates that the magnetization dynamics in stochastic MTJ driven by thermal fluctuations \cite{Bhatti17,Camsari19,Lee25} exhibit certain characteristics of chaotic behavior.  

{\it Conclusion.}
In conclusion, we theoretically analyzed nonlinear and chaotic dynamics controlled by AC and DC currents in MTJs.
The perpendicular magnetic anisotropy and the external DC magnetic field in the free layer generate a homoclinic orbit in phase space, which leads to chaotic behavior.
The AC current excites the magnetization and induces chaotic dynamics, whereas the DC current shifts the magnetization away from the homoclinic orbit, resulting in the suppression of chaos.
In order to evaluate the robustness of chaos against thermal fluctuations, we examined how a stochastic force influences the emergence of chaotic magnetization dynamics.
We found that chaos not only persists in the presence of the stochastic force, but that the stochastic force also assists to excite the magnetization and promote chaotic behavior, indicating that chaotic magnetization dynamics can be detected experimentally.
Furthermore, this result is consistent with findings for the original Duffing oscillator \cite{Strogatz24,Kovacic11}, thereby reinforcing the analogy between the two systems.
In particular, noise-induced chaos and order, which have been actively studied in random dynamical systems, are observed.
In addition, we evaluate the characteristics of chaos in magnetization dynamics driven solely by stochastic forces.
According to the Lyapunov exponent, as the variance of the stochastic force increases, the system becomes more sensitive to initial conditions, which is a feature typical of chaotic dynamics.
This indicates that the stochastic MTJ can also exhibit properties of chaotic dynamics.
This work not only advances the practical implementation of spintronic devices that utilize chaos, but also contributes to a deeper understanding of the interplay between thermal fluctuations and chaos in magnetization dynamics.

{\it Acknowledgments.}
The authors thank H. Chiba and T. Yamamoto for valuable discussions. This work was supported by JSPS KAKENHI (JP22K14591, 24K00916, and 25H02105). R. T. thanks to GP-Spin program at Tohoku University. H. M. acknowledges support from CSIS at Tohoku University.

{\it Data Availability.}

The data that support the findings of this letter are not
publicly available upon publication because it is not technically feasible and/or the cost of preparing, depositing, and hosting the data would be prohibitive within the terms of this research project. The data are available from the authors upon reasonable request.

\bibliography{ref}

\clearpage\newpage

$\ $

$\ $

\onecolumngrid  

\begin{center}
\textbf{Supplemental Materials:\\
Current-control of chaos and effects of thermal fluctuations in magnetic tunnel junctions
}
\end{center}

\section{Lyapunov exponent}
The Lyapunov exponent is widely used for analyzing nonlinear dynamics and to demonstrate sensitivity to initial conditions, the character of chaos.
The LLG equation shown in Eq.~(5) in main text can be regarded as a three-dimensional dynamical system $\bm{x} = (\theta, \varphi, z) \in \mathbb{R}^3$.
Because the phase space locally resembles a Euclidean metric and norm $||\bm{x}|| = \sqrt{\Sigma^3_{i=1} x_i^2}$, the distance between two trajectories, $\bm{x}(t)$ and $\bm{\tilde{x}}(t)$, is defined as $\delta(t) = ||\bm{x}(t) - \bm{\tilde{x}}(t)||$.
The long-term evolution of $\delta(t)$ as $\delta(0) \rightarrow 0$ characterizes the instability of the trajectories. 
The divergence rate of nearby trajectories is described by the (maximum) Lyapunov exponent
\begin{equation}
    \lambda = \lim_{t \rightarrow \infty} \frac{1}{t} \ln \frac{\delta(t)}{\delta(0)}.
\end{equation}
If $\lambda > 0$ indicates that the dynamical system is chaotic, whereas $\lambda = 0$ indicates that the dynamical system is a limit cycle. 
We compute the Lyapunov exponent using the Shimada–Nagashima method \cite{Shimada79}.

\begin{figure}[b]
\begin{centering}
\includegraphics[width=0.5\textwidth,angle=0]{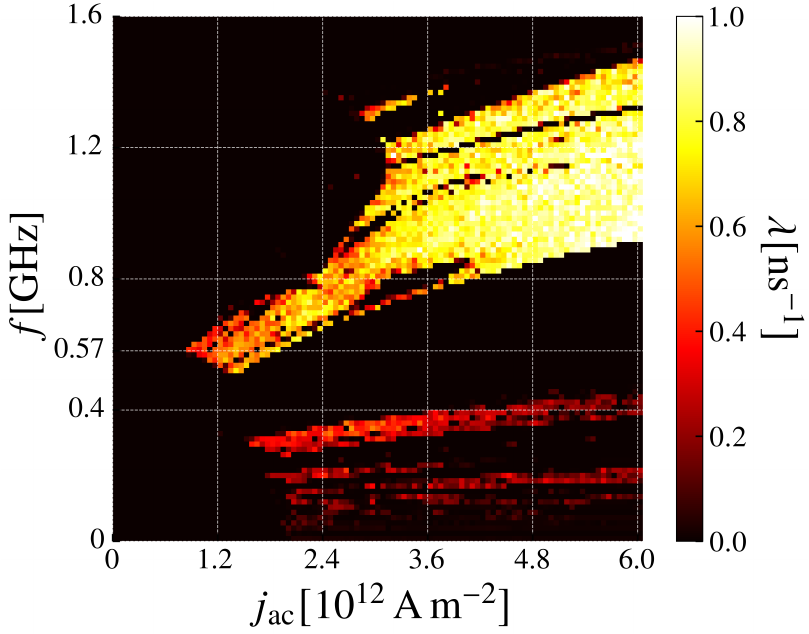} 
\par\end{centering}
\caption{
Heatmaps of the Lyapunov exponent $\lambda$ as functions of the amplitude of AC current density $j_{\mathrm{ac}}$ and the frequency of AC current $f$.}
\label{fig:Lya_omega}
\end{figure}

\section{Frequency dependence of the chaos}\label{}
Here, we examine the effect of the AC current frequency on the onset of chaos by calculating the Lyapunov exponent. 
Figure~\ref{fig:Lya_omega} presents a heatmap of the Lyapunov exponent as a function of the AC current density $j_{\mathrm{ac}}$ and its frequency $f$ without $j_{\mathrm{dc}}$ and $\sigma$.
The resonance frequency of this model can be derived from the Jacobian matrix at the potential minimum \cite{Tatsumi25} as
\begin{equation}
    f_r= \frac{\gamma}{2 \pi} \sqrt{B_K^2 - B_x^2} \approx 0.76~\mathrm{GHz} \qquad (B_x>B_K).
\end{equation}
As shown in Fig.~\ref{fig:Lya_omega}, chaos appears not only near the resonance condition but also over a wide range of frequencies $f$.
The threshold amplitude of the AC current varies significantly when the frequency $f$ changes.
In particular, the lowest threshold occurs at $f = 0.57$~GHz, which is the reason why we use $f = 0.57$~GHz in the main text.

\begin{figure*}[t]
\begin{centering}
\includegraphics[width=1.0\textwidth,angle=0]{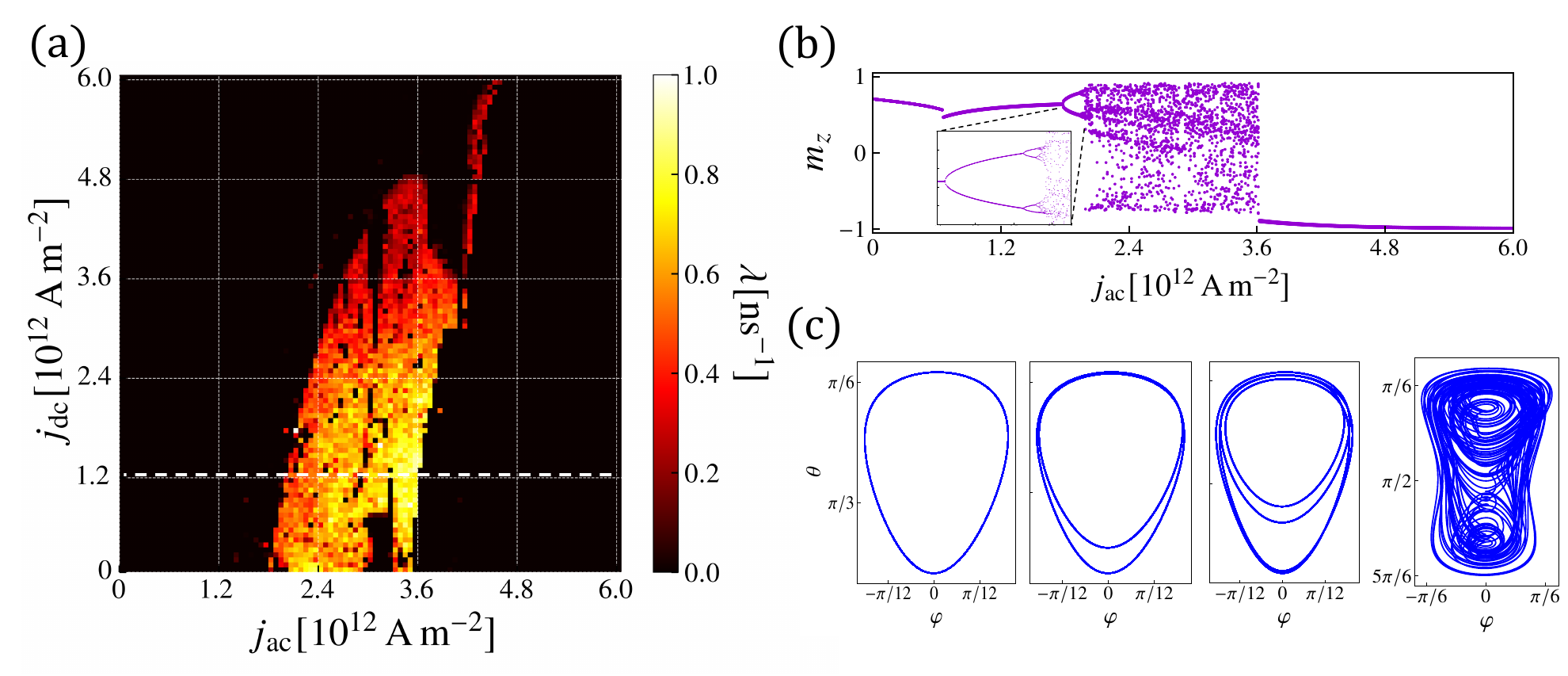} 
\par\end{centering}
\caption{
(a) Heatmaps of the Lyapunov exponent $\lambda$ as functions of the DC current density $j_\mathrm{dc}$ and the amplitude of AC current density $j_\mathrm{ac}$ under the resonance condition, where the frequency of periodic force is $0.76 \ \mathrm{GHz}$.
(b) Bifrucation diagram as a function of the amplitude of AC current density $j_\mathrm{ac}$.
(c) Magnetization trajectories for $j_\mathrm{ac}=4.0, 4.5, 4.9$ and $6.0 \times 10^{12}$~$\mathrm{Am^{-2}}$, respectively.}
\label{fig:Lya_resonance}
\end{figure*}

\section{Magnetization dynamics at the resonance}
Here, we examine the magnetization dynamics under an AC current at the resonance frequency $f_r = 0.76$ GHz by computing the Lyapunov exponent and the bifurcation diagram.
The bifurcation diagram indicates whether the dynamics are periodic or non-periodic. 
Because the system is driven by an AC current with a time period $T_r$ = $1 / f_r$, it is expected to exhibit a periodic boundary condition with the period $T_r$ in time.
Therefore, we construct a bifurcation diagram using a stroboscopic map defined as $m_y^* = {m_y(t_0 + nT_r)| n \in \mathbb{N}}$, where $t_0$ is the transient time. 
In other words, we plot the value of $m_y$ by the time period $T_r$ as a function of $j_{\mathrm{ac}}$. 
The bifurcation diagram allows us to understand how the periodicity changes as a function of the control parameter. 

Figure~\ref{fig:Lya_resonance}(a) shows a heatmap of the Lyapunov exponent as a function of the AC current density $j_{\mathrm{ac}}$ and DC current density $j_{\mathrm{dc}}$ with $f = f_r = 0.76$~GHz and $\sigma =0$.
Even under the resonance condition, the DC current increases the threshold amplitude of the AC current required for the onset of chaos.
A clear boundary between chaotic and limit-cycle regimes is observed, although no linear relationship is found between the threshold amplitudes of the AC and DC currents.
Figure~\ref{fig:Lya_resonance}(b) illustrates the bifurcation diagram of the magnetization dynamics.
The inset highlights the region where the system begins to exhibit chaotic behavior and shows that the period of the magnetization dynamics undergoes repeated doublings.
Therefore, we conclude that a period-doubling bifurcation can be observed before chaos emerges in MTJ systems.
This bifurcation is known as the famous route to chaos and is also observed in the Duffing oscillator.
The solutions in Fig.~\ref{fig:Lya_resonance}(c) show periodic orbits with the time periods $1 / f_r$, $2 / f_r$ and $4/ f_r$, as well as chaotic trajectories.
These results confirm that the period of the motion within one side of the double-well potential undergoes repeated doublings before chaos emerges.

\end{document}